\documentclass[aps,pre,onecolumn,longbibliography,superscriptaddress,nofootinbib,footnotetext]{revtex4-2}
\usepackage{bm}
\usepackage{amsmath}
\usepackage{soul}
\usepackage{color}
\usepackage{float}
\usepackage{bm,graphicx,graphics,amsmath,amssymb,epsfig,color}
\usepackage{dcolumn}
\usepackage{subcaption}

\newcommand{\beq}{\begin{equation}}
\newcommand{\eeq}{\end{equation}}
\newcommand{\beqa}{\begin{eqnarray}}
\newcommand{\eeqa}{\end{eqnarray}}

\newcommand{\bcol}[1]{{\color{black} #1}}

\begin{document}

\title{Entropy production and irreversibility in the linearized stochastic Amari
  neural model}

\author{Dario Lucente}
\affiliation{Department of Mathematics \& Physics, University of Campania “Luigi Vanvitelli”, Viale Lincoln 5, 81100 Caserta, Italy}

\author{Giacomo Gradenigo}
\affiliation{Gran Sasso Science Institute and INFN-LNGS, Viale F. Crispi 7, 67100 L'Aquila, Italy}

\author{Luca Salasnich}
\affiliation{Dipartimento di Fisica e Astronomia ``Galileo Galilei'' and INFN, Universit\`a di Padova, Via Marzolo 8, 35131 Padova, Italy}

\begin{abstract}
One among the most intriguing results coming from the application of
statistical mechanics to the study of brain is the understanding that
it, as a dynamical system, is inherently out of equilibrium. In the
realm of non-equilibrium statistical mechanics and stochastic
processes the standard observable computed to discriminate whether a
system is at equilibrium or not is the entropy produced along the
dynamics. For this reason we present here a detailed calculation of
the entropy production in the Amari model, a coarse-grained model of
the brain neural network, consisting in an integro-differential
equation for the neural activity field, when stochasticity is added to
the original dynamics. Since the way to add stochasticity is always to
some extent arbitrary, i.e., in particular for coarse-grained models,
there is no general prescription to do it, we precisely investigate
the interplay between the noise properties and the original model
features, discussing in which cases the stationary state is of thermal
equilibrium and which cases is out of equilibrium, providing explicit
and simple formulas. We also show how, following for the derivation
the particular case considered, how the entropy production rate is
related to the variation in time of the Shannon entropy of the system.
\end{abstract}


\maketitle


\section{Neural dynamics irreversibility and statistical mechanics}
\label{sec:intro0}

A possible description of large-scale brain activity is the one given
by coarse-grained models based on the dynamics of neural fields. In
such models the average activity of a large neuronal population at a
given position is encoded in the specific form of the
integro-differential equations which govern the fields
dynamics~\cite{buice2007field}. Such a description is envisaged for
understanding the behavior of large number of interacting units
without including all possible microscopic details of the real
dynamics. Since the pioneering contribution of Wilson and
Cowan~\cite{wilson1972excitatory} and of
Amari~\cite{Amari77a,amari1977dynamics}, tools from both equilibrium
and non-equilibrium statistical mechanics, with particular reference
emergence of critical
behaviours~\cite{beggs2003neuronal,buice2007field,buice2009statistical,de2014criticality,PhysRevLett.116.240601,PhysRevResearch.2.013318},
have been extensively used for characterizing neural activity. In
particular the necessity to consider {\it non-equilibrium} statistical
mechanics has been motivated by the evidence that brain activity
exhibits features of an intrinsically irreversible
process~\cite{lynn2021broken}. An issue which therefore stays at the
forefront of the interplay between theoretical physics and
neuroscience is how to use the tools of non-equilibrium statistical
mechanics to ascertain the irreversible nature of brain dynamics from
data. To this end, among other tools, the study of
fluctuation-dissipation relations, namely how internal correlations of
a given system are connected to its response to an external
perturbation, has been most widely considered. The crucial feature of
fluctuation-dissipation relations is that they take a different form
depending on whether the nature of system dynamics is irreversible or
not. This tool has been therefore considered to study the functional
relation between the evoked response to external stimuli and
steady-state correlations in brain
activity~\cite{PhysRevResearch.2.033355,PhysRevE.107.064307},
revealing its irreversible nature~\cite{PhysRevResearch.7.013301}. The
main limitation of the fluctutation-dissipation approach to the study
of irreversibility dynamics in brain activity is the necessity to
perform response experiments: the possibility to ascertain the nature
of its dynamics by simply monitoring specific indicators without the
intervention of external stimuli is therefore appealing. To this
purpose it is quite useful to study how fluctuations of relevant
observables related to brain activity are correlated in time, in
particular functions of the kind $C(t'',t')$, for which we do not need
to specify here neither the specific observable neither the structure
of the function, but we just need to know that $t'$ is the time of the
first detection of the signal and $t''$ is the time of second
detection of the signal, $t''>t'$. The role of $C(t'',t')$ is then to
tell us how the fluctuations at $t''$ are correlated to that at
$t'$. By definition the system has an equilibrium/reversible dynamics
when the time-ordered pattern of fluctuations is statistically
equivalent when the same sequence is considered from $t'$ to $t''$ or
from $t''$ to $t'$. In this situation one says that the dynamics of
the system enjoys a \emph{time-reversal} symmetry, in which case also
the correlation function $C(t'',t')$ has the same symmetry, i.e.,
$C(t'',t')=C(t',t'')$. It is then sufficient to spot out a single
observable for which the above symmetry is broken, i.e.,
$C(t'',t')\neq C(t',t'')$, to say that the time-reversal symmetry is
broken, namely the system is \emph{irreversible}. Unfortunately the
behaviour of the time-correlation functions strongly depend on the
choice of the observables and on their specific functional form, which
must often be chosen case by case. While the use of time-correlation
function presents a clear advantage, namely that one does need a
specific model for the system microscopic dynamics, it also has a
major limitation, the fact that it is not clear in all cases a priori
which is the correct observable to chose for the correlation
functions~\cite{lucente2025conceptual}. On the contrary, an
\emph{objective} and universal indicator of the lack of thermal
equilibrium is the so-called \emph{entropy production rate}, which is
zero for systems at thermal equilibrium and positive for system with
irreversible dynamics, and which can be measured directly from the
{\it "unperturbed"} dynamics, namely from the spontaneous dynamical
evolution of the system, without applying external stimuli. The only
limitation of entropy production is that in order to have a reliable
estimate of its one needs to introduce an appropriate theoretical
model of the system: it cannot be computed solely from data without
any assumptions on the underlying
model~\cite{lucente2025conceptual}. It is for this reason that, if one
wishes to study the irreversibility, and hence entropy production,
from brain data in connection to the Amari model, the first task is to
study the explicit formula for entropy production which can be
obtained for the case at hand. In particular, since the original
equations of the model are deterministic, a necessary technical step
in order to use the standard formalism to compute entropy production,
without consider the formalism proper to deterministic
systems~\cite{ruelle1996positivity}, is to randomize the dynamics. The
task of the present discussion is precisely to show how different ways
of dynamics randimization connect to the intrinsic properties of the
models yielding different results for the entropy production.\\

The discussion of the paper is structured as follows: in
Sec.~\ref{sec:entropy-intro} we recall from the realm of
non-equilibrium statistical mechanics and stochastic thermodynamics
the modern definition of entropy production in stochastic systems,
first provided in a famous paper of Lebowitz and
Spohn~\cite{lebowitz1999gallavotti}, mentioning also that it is
related to the Shannon entropy of the system of interest. The purpose
of this section is also to motivate the choice of entropy production
rather other observables to characterize the degree of irreversibility
in the system.  Since the tools introduced in
Sec.~\ref{sec:entropy-intro} apply specifically to stochastic systems,
in Sec.~\ref{sec:stoch-amari-eq} we introduce a linearized version of
the Amari equation, proposing a standard protocol to randomize it and
commenting which is the simplest choice for model properties which
leads to an equilibrium state with reversible dynamics, where the
entropy production is zero. Then in Sec.~\ref{sec:entropy-prod-gen} we
present our first original result: the complete calculation of the
entropy production as in indicator of irreversibility in the
Lebowitz-Spohn approach, where the probability of trajectories forward
in time is compared with the probability of trajectories backward in
time. The resulting expression clarifies the peculiar interplay
between the Amari model properties and the properties of the
superimposed stochasticity in determining the irreversible or
irreversible nature of the dynamics. Sec.~\ref{sec:Shannon} is then
dedicated to show how the same expression obtained in
Sec.~\ref{sec:entropy-prod-gen} from the Lebowitz-Spohn formalism can
be obtained by different means studying the rate of variation in time
of the Shannon entropy, another concept typical of stochastic
thermodynamics. The peculiarity of this section is the discussion in
terms of a functional formalism for both the Shannon entropy and the
Fokker-Planck equation governing the stochastic dynamics of the neural
activity field, which is not so commont in the
literature. Furthermore, by comparing the Lebowitz-Spohn entropy
production with the variation of Shannon entropy, this section is
useful to grasp a better physical insight in the object of
study. Sec.~\ref{sec:EntropyFourier} is then dedicated to show how how
a simple explicit expression of the entropy production can be computed
when assuming translational invariance in the space of coordinates for
the original equation, making transparent the role played by the
symmetry properties of the linearized Amari equation force. Finally in
Sec.~\ref{sec:conclusion} we turn to conclusions.

\section{Entropy production in non-equilibrium statistical mechanics}
\label{sec:entropy-intro}

As anticipated in the previous section, an important concept which
neuroscience, in particular the study of brain activity, has borrowed
from non-equilibrium statistical mechanics, is the one of entropy
production~\cite{kringelbach2024thermodynamics,nartallo2025nonequilibrium},
a measurable quantity which characterizes the degree of
irreversibility of brain dynamics. For instance, entropy production
has been related to the hierarchical and asymmetrical structure of
brain connections~\cite{PhysRevE.110.034313,nartallo2025multilevel,geli2025non},
with techniques successfully applied also to the analysis of
experimental signals~\cite{PhysRevResearch.2.033355,lynn2021broken}. It has also
been shown that different estimates of entropy production from
empirical datasets reveal a correlation to consciousness level and to
the activity recorded in subjects performing different
tasks~\cite{lynn2021broken,PhysRevE.107.024121,PhysRevResearch.7.013301,PhysRevE.110.034313}. Moreover,
an interesting decomposition of entropy production in oscillatory
modes has been applied to study how the brain rhythms contribute to
dissipation and information
processing~\cite{sekizawa2024decomposing}. Let us therefore spend some
words on the definition and meaning of entropy production. The fact
that entropy, which is a state function of a macroscopic system, grows
along an irreversible transformation is a well known fact from the
19th century. An account of how to related quantitatively the entropy
constantly produced along the irreversible dynamics of a stationary
non-equilibrium state with macroscopic transport coefficients can be
for instance found in the historical book of De Groot and
Mazur~\cite{de2013non}, which yields an overview on the development of
the subject in the second half of the 20th century. How the entropy
produced macroscopically along stationary irreversible dynamics can be
related to the probability of microscopic trajectories is a much more
recent achievement. The standard way to compute entropy production in
all stochastic process has been settled by the seminal paper of
Lebowitz and Spohn in 1999~\cite{lebowitz1999gallavotti}. This work
was proposed to extend the definition of the Gallavotti-Cohen symmetry
for the probability of stationary non-equilibrium fluctuations,
initially conceived for deterministic dissipative system, to systems
which can be characterized in terms of a Markov chain. While in
deterministic systems the entropy produced macroscopically can be
related, within the framework of a refined mathematical analysis, to
the rate of contraction of phase space~\cite{ruelle1996positivity},
the work of Lebowitz and Spohn has been the first to provide an
explicit formula to relate the same quantity to the probability of
system trajectories. And, let us add, it is precisely the overwhelming
simplicity of computing the entropy production in presence of
stochasticity that motivated our present work, where we introduce the
stochastic version of the Amari equation and compute explicitly the
related entropy production rate. According to the Lebowitz-Spohn
formula, if one considers the time interval $[0,t]$ and denotes the
sequence of configurations visited by the system in this interval as
$\Omega_0^t$, which is usually denoted as {\it forward} trajectory,
and the reversed sequence of configurations as
$\overline{\Omega}_0^t$, which is usually denoted as the
\emph{backward} trajectory, the entropy produced along the time span
$[0,t]$ reads
as~\cite{lucente2025conceptual,sarracino2025nonequilibrium}
\begin{align}
\Sigma(t) = \log\left( \frac{\mathcal{P}[\Omega_0^t]}{\mathcal{P}[\overline{\Omega}_0^t]} \right)
\end{align}
One of the most interesting properties of the entropy production is
that in general, i.e., systems with finite-term memory, it can be
shown that $\Sigma(t)$, which depends on the stochasticity of the
dynamics, enjoys a self-averaging property~\cite{nardini2017entropy},
namely in the large-system size its probability distribution naturally
concentrates around the average, that is:
\begin{align}
\lim_{t\rightarrow\infty}  \frac{\Sigma(t)}{t} = \lim_{t\rightarrow\infty}\frac{\langle \Sigma(t) \rangle}{t} ,
\label{eq:ent-prod-rate}
\end{align}
where, in the above expression, the average is taken with respect to
the probability of forward trajectories and we have divided by $t$ in
order to have a finite quantity, since $\Sigma(t)$ is by definition
extensive in $t$. Therefore, since at large time we have $\Sigma(t)
\approx \langle \Sigma(t) \rangle$, the properties of the typical
value of entropy production for $t\gg1$ are those of the mean, which
reads as
\begin{align}
\langle \Sigma(t) \rangle = \int \mathcal{D}\Omega_0^t~\mathcal{P}(\Omega_0^t)~\log\left( \frac{\mathcal{P}[\Omega_0^t]}{\mathcal{P}[\overline{\Omega}_0^t]}\right).
\label{eq:ent-prod-av}
\end{align}
What is nice about the expression in Eq.~\eqref{eq:ent-prod-av} is
that it takes explicitly the form of a Kullback-Leibler
divergence. This is particularly interesting in order to motivate the
choice of entropy production rather than correlation functions in
order to assess the degree of irreversibility of the dynamics. As the
distance between two probability distributions, the Kullback-Leibler
divergence does not depend on the choice of variables, as long as the
probabilities $\mathcal{P}(\Omega_0^t)$ and
$\mathcal{P}(\overline{\Omega}_0^t)$. This gives to the entropy
production a great degree of universality: it does not depend on the
variables we choose to represent the system trajectories, as long as
we are able to determine with precision their probability. In
addition, if some degrees of freedom are integrated out, the
Kullback-Leibler divergence cannot increase, meaning that the entropy
production rate decreases under
coarse-graining~\cite{lucente2025conceptual}. On the other hand, the
"limit" of this approach is that we need a precise mathematical
modeling for the dynamics of our system, otherwise it is quite
unlikely to access the correct probability distributions. The general
manipulations done in the literature starting from the definition of
entropy production in Eq.~\eqref{eq:ent-prod-av} are usually aimed at
writing the stationary entropy production rate in terms of the
appropriate combination of correlation functions, a task which is
accomplished by computing, in any specific case, the quantity
\begin{align}
\sigma = 
\lim_{t\rightarrow\infty} \frac{\Sigma(t)}{t} ~~=~~ \text{stationary correlators} .
\end{align}
The purpose of the work presented here is precisely to illustrate how
by adding stochasticity to the originally deterministic Amari model it
is possible to exploit the entropy production formulae which are
widespread in the contemporary
literature~\cite{ruelle1996positivity,lebowitz1999gallavotti,Sekimoto2010,seifert2012stochastic}. In
addition to that, we will show how the same expression for the entropy
production rate can be obtained, partly by going through known
formulae~\cite{Sekimoto2010,peliti2021stochastic,sarracino2025nonequilibrium},
also starting from the Shannon entropy formula from stochastic
thermodynamics. This alternative derivation of the same formulae will
clarify the whole picture, connecting the entropy production with the
time variation of the stationary entropy of the system. Let us assume
that all variables needed to specify the state of the system are
indicated with $X$ and that, their time-dependent probability
distribution $P(X,t)$ is known. In this case, the Shannon entropy of
the system reads as
\begin{align}
S_{\text{sys}}(t) = - \int \mathcal{D}X~P(X,t)~\log[P(X,t)],
\label{eq:sh-1}
\end{align}
where we have used the symbol $\mathcal{D}X$ to denote functional
integration over a possibly multidimensional space. By simple formal
manipulations, also using the Fokker-Planck equation of which $P(X,t)$
is a solution, it is a standard procedure to show that the
time-derivative of the system entropy, defined as the Shannon entropy
in Eq.~\eqref{eq:sh-1}, splits into the difference of two
contributions, which are usually interpreted respectively as the
entropy of the universe, $S_{\text{tot}}$, and the entropy of the
reservoir coupled to the system, $S_{\text{res}}$:
\begin{align}
\dot{S}_{\text{sys}}(t) = \dot{S}_{\text{tot}}(t) - \dot{S}_{\text{res}}(t).
\end{align}
In a stationary state the probability distribution $P(X,t)$ is
stationary, so that also the entropy of the system is constant and
$\dot{S}_{\text{sys}}(t)=0$. In this case the rate of increase of the
entropy of the universe and the entropy of the reservoir are
identical, $\dot{S}_{\text{tot}}(t)=\dot{S}_{\text{res}}(t)$, and we
will show that they also correspond to the system's entropy production
rate as computed from a Lebowitz-Spohn like functional. So far, we
kept the discussion general in order presents all the aspects of
entropy production which are non-specific of a particular model. The
next section will be devoted to the presentation of the Amari model,
which we have chosen as a benchmark to study entropy production in a
model for brain dynamics. \\

\section{Linearized stochastic Amari model: equilibrium properties}
\label{sec:stoch-amari-eq}

In the previous section we have introduced the concept of entropy
production rate as an indicator for the degree of irreversibility in
system dynamics. We have also mentioned how, due to the Lebowitz-Spohn
generalization of the Gallavotti-Cohen results, the calculation of
entropy production is much easier in system with stochasticity. It is
for this reason that, elaborating on its original deterministic
version, we have added stochasticity to the Amari model for neural
dynamics. Before presenting the calculation of the entropy production
in this model, let us introduce the model and motivate its choice.
The stochastic Amari model is an integro-differential stochastic
equation which models the neural activity of the brain by means of a
local activation field $u({\bf x},t)$ defined in a spatio-temporal
domain, ${\bf x} \in \Omega \subset \mathbb{R}^d$ and $t\in
\mathbb{R}$ \bcol{(we also assume $\Omega$ to be a boundary free
  domain)}:
\beq
\frac{\partial}{\partial t} u({\bf x},t) = -\frac{1}{\tau}u({\bf x},t)
+ \frac{1}{\tau}\int_{\Omega} w({\bf x},{\bf y}) \, 
f[u({\bf y},t)] \, d{\bf y}  + \xi({\bf x},t) \; ,
\label{eq:stochastic-Amari}
\eeq
where $\tau$ is the relaxation time and the two-point function $w({\bf
  x},{\bf y})$, which models the transmission of impulses from one
region to another of the brain, is usually known as the synaptic
weight. In Eq.~\eqref{eq:stochastic-Amari} then appears $f[u({\bf
    x},t)]$, the activation function, which is expected saturate to a
costant value when the local field exceeds a certain threshold
$u^*$. Usually the activation function is modelled as a sigmoid:
\beq
f[u] = {1\over e^{\beta (u^* - u)} + 1} \;
\label{fermi}
\eeq
with gain $\beta >0$ and threshold $u^*>0$. For the present discussion
we will consider $\xi({\bf x},t)$ to be a Gaussian field
delta-correlated in time, i.e., a white noise. Beside the correlation
in the time domain, the noise field is also characterized by its
spatial covariance:
\begin{align}
\langle \xi({\bf x},t) \xi({\bf y},t')\rangle = \gamma({\bf x},{\bf y})~\delta(t-t'),
\end{align}
where, in order to a have a consistent definition of the noise, its
covariance $\gamma({\bf x},{\bf y})$ must be a symmetric function of
its coordinates:
\begin{align}
\gamma({\bf x},{\bf y})=\gamma({\bf y},{\bf x}).
\label{eq:gammacose}
\end{align}
For simplicity, here we only consider the case of an additive
noise. Thus, Eq.~\eqref{eq:stochastic-Amari} does not suffer from
different physical interpretation resulting from different
discretization schemes~\cite{sokolov2010ito}. This choice allow us to
focus on the role of synaptic couplings and noise correlations in
sustaining irreversible stationary states without facing all the
difficulties encountered when multiplicative noise, with possibly
time-varying power-law behavior, is taken into
account~\cite{cherstvy2015ergodicity}. We have chosen a stochastic
version of the Amari neural-field equation because it provides a
minimal, spatially continuous mean-field description that directly
links macroscopic cortical activity to physiologically interpretable
ingredients: the synaptic coupling kernel $w({\bf x},{\bf y})$ encodes
anatomical connectivity, the nonlinear gain function $f[u({\bf x},t)]$
represents neuronal input–output properties, and explicit temporal
decay modeled by the term $-\tau^{-1}u({\bf x},t)$ represents
membrane/leak dynamics. The deterministic Amari field $u({\bf x},t)$
was originally introduced as a canonical model for pattern formation
and spatially structured activity in cortex (bumps, waves, spatial
patterns), as discussed in Chap. 1 of Ref.
\cite{coombes2014neural}. The addition of a stochastic forcing to this
framework is not only a technical trick to allow for a much easier
calculation of the entropy production but arise also a natural and
well-studied extension because real cortical tissue is subject to
numerous sources of variability (synaptic noise, finite-size effects,
background inputs) whose principal effects are captured at the
continuum level by noise terms (see Chap. 9 of
Ref. \cite{coombes2014neural}). In addition to that, stochastic
neural-field equations have been used to analyze phenomena such as
noise-induced wandering of stationary bumps, diffusion of wave
positions, variability of stimulus tuning, and noise-driven
bifurcations of spatial patterns — phenomena directly relevant to
working memory, perceptual switching, and cortical variability
observed in
experiments~\cite{bressloff2011spatiotemporal,coombes2014neural,amari1977dynamics}. From
the mathematical perspective, the stochastic Amari model admits a
rigorous probabilistic treatment (well-posedness, invariant measures,
ergodicity) under reasonable assumptions on the synaptic kernel
$w({\bf x},{\bf y})$ and noise correlations $\gamma({\bf x},{\bf y})$,
which both justifies the analytical approach taken here. This
combination of physiological interpretability, direct connection to
experimentally observed noisy dynamics, and a mature mathematical
literature motivated our choice of this model for randomization.\\

Since the Amari model is originally defined as a deterministic model,
there is a priori some degree of arbitrariness in the choice of the
noise properties. In fact, whether the system attains or not a
stationary state which can be deemed as equilibrium depends, as we are
going to show, from the interplay between the properties of noise and
those of the local force acting on the neural activity field. Here we
start by showing which choice of noise yields equilibrium properties,
leaving a more detailed study of the general case to the next
section. As a first step in this direction it is convenient to
linearize the Amari equation around a stationary solution. One can
therefore denote as $\eta({\bf x},t)$ the fluctuations around the
homogeneous solution $u({\bf x},t)=u_0$:
\beq
\eta({\bf x},t) = u({\bf x},t) -u_0 \;  \quad\quad\quad \left|\frac{u({\bf x},t)-u_0 }{u_0}\right|\ll 1 .
\label{eq:field-fluct}
\eeq
We consider then the expansion to linear order in $\eta({\bf x},t)$ of
the non-linear activation function $f[u({\bf x},t)]$:
\beq f[u({\bf x},t)] = f[u_0] + f'[u_0] \eta({\bf x},t) + ... \; 
\label{eq:linear-activ}
\eeq
The homogeneous solution $u_0$ is the one obtained by plugging
$\eta({\bf x},t) = u_0 + u({\bf x},t)$ into
Eq.\eqref{eq:stochastic-Amari}, expanding and solving for $u_0$ after
having set $\dot{\eta}({\bf x},t)=0$ and $\xi({\bf x},t)=0$:
\beq
0 = -u_0 + \tilde{w} \, f[u_0] \;  
\eeq
where
\beq 
\tilde{w} = \int_{\Omega} w({\bf x},{\bf y}) \, d{\bf y}. 
\label{eq:wtilde}
\eeq
The claim that the integral in the right-hand side of
Eq.~\eqref{eq:wtilde} yields a constant value $\tilde{w}$ independent
from the coordinate ${\bf x}$ can be simply justified under the
\bcol{hypothesis} that all points in the brain are equivalent from the
perspective of the connectivity with the rest of the system, which is
quite reasonable if one considers that all models for neural networks
in the brain~\cite{PhysRevLett.61.259,PhysRevE.98.062120} are usually
dense networks.  For a given choice of $\tilde{w}$, the stationary
homogeneous solution $u_0$ can be simply obtained from the following
algebraic non-linear equation
\beq
f[u_0] = {u_0 \over \tilde{w}} \; .
\eeq
By plugging the expansion of Eq.\eqref{eq:field-fluct} into the
stochastic Amari equation and truncating to the linear order we get
\beq \frac{\partial}{\partial t} \eta({\bf x},t) = -\frac{1}{\tau}\eta({\bf
  x},t) + \frac{f'[u_0]}{\tau} \int_{\Omega} w({\bf x},{\bf y}) \eta({\bf y},t)
d{\bf y} +  \xi({\bf x},t) \; ,
\label{slae}
\eeq
which is the stochastic linear Amari equation for the (small)
fluctuations $\eta({\bf x},t)$ around the homogeneous solution
$u_0$. The last equation can be rewritten in a more compact form as
follows:
\begin{align}
\dot{\eta}({\bf x},t) = \Lambda[\eta]({\bf x},t) + \xi({\bf x},t),
\label{eq:Ama-compact}
\end{align}
where $\Lambda[\eta]({\bf x},t)$ is an integral operator, acting on the field $\eta({\bf x},t)$, defined by its kernel
$\lambda({\bf x},{\bf y})$
\begin{align}
  \Lambda[\eta]({\bf x},t) & = \int d{\bf y}~\lambda({\bf x},{\bf y})~\eta({\bf y},t) \nonumber \\
  \lambda({\bf x},{\bf y}) & = - \frac{1}{\tau} \delta^{(d)}({\bf x},{\bf y}) +
  \frac{f'[u_0]}{\tau} w({\bf x},{\bf y})  
\end{align}
One of the goals of the following discussion is to study how the
presence of a thermodynamic equilibrium depends on the properties of
$\lambda({\bf x},{\bf y})$. Clearly, the operator $\Lambda$ must be
bounded and negative define to guarantee that
Eq.~\eqref{eq:Ama-compact} is well-defined. For instance, a sufficient
condition is that $\lambda\in L_2(\Omega)$, being $L_2(\Omega)$ the
space of square-integrable function on $\Omega$. At the same time, the
noise correlation kernel $\gamma$ must be positive definite and, for
simplicity, we also assume it is invertible.
It is well known that when the synaptic weight $w({\bf x},{\bf y})$ is
a symmetric function of the coordinates and $\Lambda$ is a
self-adjoint operator. Therefore, one can introduce the potential:
\begin{align}
U[\eta] = - \bcol{1\over 2} \int d{\bf x}~d{\bf y}~\eta({\bf x})\lambda({\bf x},{\bf y}) \eta({\bf y})
\end{align}
and write the Langevin equation as
\begin{align}
\dot{\eta}({\bf x},t) = - \frac{\delta U[\eta]}{\delta \eta({\bf x},t)} + \xi({\bf x},t). 
\end{align}
In this case it is sufficient to have a noise which is
delta-correlated in both space and time, i.e., a delta-correlated spatial kernel
\begin{align}
\gamma({\bf x},{\bf y}) = T~\delta({\bf x}-{\bf y})
\end{align}
in order to have a Boltzmann-like an equilibrium distribution \bcol{(here the Boltzmann constant $k_B$ is set equal to $1$)}
\begin{align}
P_{\text{eq}}[\eta] \propto \exp\left( - U[\eta]/T \right). 
\label{eq:stationary-equilibrium-pdf}
\end{align}
We will demonstrate in the following sections that there is a more
general condition, involving also the properties of the noise kernel,
which guarantees the presence of thermodynamic equilibrium.

\section{Entropy production from the Lebowitz-Spohn formula}
\label{sec:entropy-prod-gen}

After the historical introduction on entropy production presented in  Sec.~\ref{sec:entropy-intro}, here we show how to compute it exactly in the stochastic Amari model. Let us just notice that, at variance with the original entropy production formula given in the Lebowitz-Spohn paper for Markov chains~\cite{lebowitz1999gallavotti}, where trajectories are written in terms of transition rates, the stochastic Amari model discussed here is a continuous time stochastic processes, so that we will need the functional Onsager-Machlup formalism to write path probabilities~\cite{onsager1953fluctuations,machlup1953fluctuations,sarracino2025nonequilibrium}. First of all, we have to provide an explicit definition for trajectories, which have been referred to only in an abstract way in Sec.~\ref{sec:entropy-intro}. In the present discussion the {\it forward} trajectory is represented by the collection of the field $\eta({\bf x},t)$ configurations, ordered cronologically
\begin{align}
\Omega_0^t = \left\lbrace \eta({\bf x},s) ~|~ s \in [0,t] \right\rbrace, 
\end{align}
while the symbol $\overline{\Omega}_0^t$ denotes the {\it backward} trajectory
\begin{align}
\overline{\Omega}_0^t = \left\lbrace \overline{\eta({\bf x},s)} ~|~ s \in [0,t] \right\rbrace.
\end{align}
where the overline denotes the sequence of configurations defined as follows 
\begin{align}
  \overline{\eta({\bf x},s)} &= \eta({\bf x},t-s)~~~s~\in~[0,t] \label{eq:eta-back} \\
  \frac{d}{ds}\overline{\eta({\bf x},s)} &= - \frac{d}{dt}\eta({\bf x},t-s)~~~s~\in~[0,t] \label{eq:eta-dot-back}. 
\end{align}
The apparently ambiguous notation for the derivative on the right-hand
term of Eq.~\eqref{eq:eta-dot-back} must be interpreted as:
\begin{align}
\frac{d}{dt}\eta({\bf x},t-s) = \frac{d}{du}\eta({\bf x},u)\bigg|_{u=t-s}\,.
\end{align}
The task is then to compute
\begin{align}
\Sigma(t) = \log\left( \frac{\mathcal{P}[\Omega_0^t]}{\mathcal{P}[\overline{\Omega}_0^t]} \right),
\end{align}
for which we need the expression of $\mathcal{P}[\Omega_0^t]$ and of
$\mathcal{P}[\overline{\Omega}_0^t]$.  According to the
Onsager-Machlup formula, for which the derivation is standard in the
case of additive noise, see for
instance~\cite{peliti2021stochastic,sarracino2025nonequilibrium}, the
probability of forward and backward trajectories can be respectively
written as:
\begin{align}
& \mathcal{P}[\Omega_0^t] \propto \exp\left\lbrace - \frac{1}{2} \int_0^t ds
\int d{\bf x}~d{\bf y} \left[ \dot{\eta}({\bf x},s)-\Lambda[\eta]({\bf x},s)\right]
\gamma({\bf x},{\bf y})^{-1} \left[\dot{\eta}({\bf y},s)-\Lambda[\eta]({\bf y},s)\right] \right\rbrace, \nonumber \\
& \mathcal{P}[\overline{\Omega}_0^t] \propto  \nonumber \\ 
& \exp\bigg\lbrace - \frac{1}{2} \int_0^t ds
\int d{\bf x}~d{\bf y} \left[ - \dot{\eta}({\bf x},t-s)-\Lambda[\eta]({\bf x},t-s)\right]
\gamma({\bf x},{\bf y})^{-1} \left[-\dot{\eta}({\bf y},t-s)-\Lambda[\eta]({\bf y},t-s)\right] \bigg\rbrace,
\end{align}
From the above formulas, a straightforward algebraic manipulation yields the total entropy production, which takes the form of an integral over the elapsed time:
\begin{align}
\Sigma(t) &= \int_0^t ds \int d{\bf x}~d{\bf y}~\left[\dot{\eta}({\bf x},s)\gamma({\bf x},{\bf y})^{-1}\Lambda[\eta]({\bf y},s)+\Lambda[\eta]({\bf x},t-s)\gamma({\bf x},{\bf y})^{-1}\dot{\eta}({\bf y},t-s)\right] \nonumber \\
&= 
2 \int_0^t ds \int d{\bf x}~d{\bf y}~\dot{\eta}({\bf x},s)\gamma({\bf x},{\bf y})^{-1}\Lambda[\eta]({\bf y},s)
\label{eq:sdot-general}
\end{align}
where the last integral has to be performed with the Stratonovich prescription. This is not related to the discretization scheme adopted in the definition of the stochastic differential equation and can be derived by discretizing the dynamics in time intervals $\Delta t$, writing the path probability $\mathcal{P}[\Omega_0^t]$ and formally perform the limit $\Delta t\to 0$, as originally conceived by Onsager~\cite{onsager1953fluctuations,machlup1953fluctuations}.
This is the most general expression of the entropy production in the
Amari model, which relies on no other assumption than the
linearization of the equations of motion. No translational invariance in space
or symmetry under the exchange of coordinates has been made for the synaptic weight function $w({\bf x},{\bf y})$. It is then customary to consider the entropy production per unit time in the large-time limit, i.e., $\sigma(t) = \Sigma(t)/t$, which reads as a time average over and where, it the dynamics relax to a stationary state, it is possibile to replace this average with the average over the stationary distribution, thus obtaining and expression in terms of stationary correlation functions:
\begin{align}
  \sigma &= \lim_{t\rightarrow\infty}\frac{2}{t}\int d{\bf x}~d{\bf y}~d{\bf z}~\gamma({\bf x},{\bf y})^{-1}\lambda({\bf y},{\bf z})\dot{\eta}({\bf x},s)\eta({\bf z},s) \nonumber\\
& = 2 \int d{\bf x}~d{\bf y}~d{\bf z}~\gamma({\bf x},{\bf y})^{-1}\lambda({\bf y},{\bf z})\langle\dot{\eta}({\bf x})\eta({\bf z})\rangle
  \label{eq:sdot-general}
\end{align} 
What is now crucial in the manipulation of Eq.~\eqref{eq:sdot-general} is to assume the Stratonovich convention for the definition of stochastic integrals. By replacing $\dot{\eta}({\bf x})$ with its expression according to the Langevin equation, Eq.~\eqref{eq:Ama-compact}, one gets:
\begin{align}
\sigma = 2\int d{\bf x}~d{\bf y}~d{\bf z}~d{\bf z'}~\gamma({\bf x},{\bf y})^{-1}\lambda({\bf y},{\bf z})\left[\lambda({\bf x},{\bf z'})\langle\eta({\bf z'})\eta({\bf z})\rangle+\delta({\bf x}-{\bf z'})\langle\xi({\bf z'})\eta({\bf z})\rangle\right]
\label{eq:sdot-general-1}
\end{align}
The above formula for the entropy production rate $\sigma$ consists of two terms: the first one involves only $c({\bf z'},{\bf z}) = \langle\eta({\bf z'})\eta({\bf z})\rangle$, the equal-time two-point correlation function of the field. The other term depends on the {\it equal-time} correlation between the noise and the field, which is also non-trivial due to the Stratonovich prescription. Therefore, Eq.~\eqref{eq:sdot-general-1} can be rewritten as
\begin{align}
  \sigma & = 2\int d{\bf x}~d{\bf y}~d{\bf z}~d{\bf z'}~\gamma({\bf x},{\bf y})^{-1}\lambda({\bf y},{\bf z})\left[\lambda({\bf x},{\bf z'})c({\bf z'},{\bf z})+\delta({\bf x}-{\bf z'})\frac{1}{2}\gamma({\bf z'},{\bf z})\right]\nonumber\\
  & = 2\int d{\bf x}~d{\bf y}~d{\bf z}~d{\bf z'}~\gamma({\bf x},{\bf y})^{-1}\lambda({\bf y},{\bf z})\lambda({\bf x},{\bf z'})c({\bf z'},{\bf z})+\int d{\bf x}~d{\bf y}~d{\bf z}~\gamma({\bf x},{\bf y})^{-1}\lambda({\bf y},{\bf z})\gamma({\bf x},{\bf z})\nonumber\\
  & = 2\int d{\bf x}~d{\bf y}~d{\bf z}~d{\bf z'}~\lambda^T({\bf z'},{\bf x})\gamma({\bf x},{\bf y})^{-1}\lambda({\bf y},{\bf z})c({\bf z},{\bf z'})+\int d{\bf z}~\lambda({\bf z},{\bf z})\,.
  \label{eq:sdot-general_final}
  \end{align}
Let us now rewrite the expression in Eq.~\eqref{eq:sdot-general_final} in a more transparent form. In order to do that, let us first define the trace of a generic linear operator 
\begin{align} 
    \mathcal{K}[\eta]=\int d{\bf y} ~k({\bf x},{\bf y})\eta({\bf y})
\end{align} 
as
\begin{align}
    \text{Tr}[\mathcal{K}]=\int d{\bf x}~k({\bf x},{\bf x})\,.
\end{align}
By considering also the Lyapunov equation, which is a fundamental equation that must be fulfilled in order to guarantee the existence of a stationary state in linear system~\cite{gardiner2004handbook},
\begin{align}
    \int d{\bf z}~\left[\lambda ({\bf x},{\bf z}) c({\bf z},{\bf y})+c({\bf x},{\bf z})\lambda^T({\bf z},{\bf y})\right]&=-\gamma({\bf x},{\bf y})\,\nonumber \\ 
    \Lambda\circ C + (\Lambda\circ C)^T &= - \Gamma
\end{align}
it is possible to plug it into the expression of the entropy production rate in Eq.~\eqref{eq:sdot-general_final} then obtaining:
  \begin{align}
  \sigma & = 2\int d{\bf x}~d{\bf y}~d{\bf z}~d{\bf z'}~\lambda^T({\bf z'},{\bf x})\gamma({\bf x},{\bf y})^{-1}\lambda({\bf y},{\bf z})c({\bf z},{\bf z'})+\int d{\bf z}~\lambda({\bf z},{\bf z})\nonumber\\
  & = \text{Tr}\left[2\Lambda^T\circ \Gamma^{-1}\circ\Lambda \circ C+\Lambda\right]\nonumber\\
  & = \text{Tr}\left[\Lambda^T\circ \Gamma^{-1}\circ\Lambda \circ C\right]+\text{Tr}\left[ \Lambda^T\circ \Gamma^{-1}\circ\Lambda \circ C + \Lambda\right]\nonumber\\
  & =\text{Tr}\left[\Lambda^T\circ \Gamma^{-1}\circ\Lambda \circ C\right]+\text{Tr}\left[\Lambda^T \circ \Gamma^{-1}\circ (-C\circ  \Lambda^T-\Gamma)+\Lambda\right]\nonumber\\
  & =\text{Tr}\left[\Lambda^T\circ    \Gamma^{-1}\circ\Lambda \circ C\right]-\text{Tr}\left[\Lambda^T \circ \Gamma^{-1} \circ C \circ  \Lambda^T\right]+\text{Tr}\left[\Lambda-\Lambda^T\right]\nonumber\\
  & = \text{Tr}\left[(\Lambda^T \circ\Gamma^{-1}-\Gamma^{-1}\circ\Lambda)\circ\Lambda\circ C\right] 
\label{eq:sdot-general_final_trace}
\end{align} 
The above expression of entropy production in Eq.~\eqref{eq:sdot-general_final_trace} is the central result of our paper. We will show how it can be written in a more explicit and insightful form when also translational invariance is considered in Sec.~\ref{sec:EntropyFourier}. In order to clarify its derivation, let us remark that from the first to the last line of Eq.~\eqref{eq:sdot-general_final_trace} we took advantage of the properties of the trace and of several identities which we are going to list right now. First of all, from the second to the third line of Eq.~\eqref{eq:sdot-general_final_trace} we used the Lyapunov stability condition in the following form:
\begin{align}
\Lambda\circ C + (\Lambda\circ C)^T = - \Gamma~~~\Longrightarrow~~~\Lambda\circ C = -C\circ\Lambda^T - \Gamma
\end{align}
In the passages of Eq.~\eqref{eq:sdot-general_final_trace} we made also use of the cyclic permutation of the trace, of the fact that for any matrix $A$ we have $\text{Tr}[A]=\text{Tr}[A^T]$ and of the fact that $\Gamma$ and $C$, as well as their inverse, are symmetric. The above properties allowed us to write
\begin{align}
\text{Tr}\left[ \Lambda^T\circ\Gamma^{-1}\circ C \circ \Lambda^T\right] &= \text{Tr}\left[ (\Lambda \circ C \circ \Gamma^{-1}\circ \Lambda)^T\right] 
\nonumber\\
&= \text{Tr}\left[ \Lambda \circ C \circ \Gamma^{-1}\circ \Lambda\right] = \text{Tr}\left[ \Gamma^{-1}\circ\Lambda\circ\Lambda\circ C\right],
\end{align}
which, considering that the trace of an antisymmetric matrix is zero, $\text{Tr}\left[ \Lambda - \Lambda^T\right]=0$, led us to the last line of Eq.~\eqref{eq:sdot-general_final_trace}. 
The expression in the last line of Eq.~\eqref{eq:sdot-general_final_trace} makes clear that the condition 
\begin{align}
\Lambda^T\circ\Gamma^{-1} = \Gamma^{-1}\circ \Lambda,
\label{eq:zero-sigma-cond}
\end{align}
is the one necessary to have zero entropy production rate and is therefore as a consequence the one necessary to have thermodynamic equilibrium. In order to rewrite the identity in Eq.~\eqref{eq:zero-sigma-cond} in a more transparent way let us multiply both terms to the left and to the right by the operator $\Gamma$, e.g., $\Lambda^T\circ\Gamma^{-1} \rightarrow \Gamma\circ\Lambda^T\circ\Gamma^{-1}\circ\Gamma$, which, by also recalling that $\Gamma = \Gamma^T$, yields 
\begin{align}
\Gamma^T\circ\Lambda^T = \Lambda\circ\Gamma,
\label{eq:zero-sigma-cond-1}
\end{align}
which is the general "equilibrium" condition we were seeking for. A more transparent expression can be obtained by writing explicitly the action of the operators in Eq.~\eqref{eq:zero-sigma-cond} on the activity fields, for instance by writing $\Lambda \circ \Gamma[\eta]({\bf x})$ as
\begin{align}
\Lambda \circ \Gamma[\eta]({\bf x}) = \int d{\bf z}~d{\bf y}~\lambda({\bf x},{\bf z})~\gamma({\bf z},{\bf y})~\eta({\bf y}).
\end{align}
The identity in Eq.~\eqref{eq:zero-sigma-cond-1} can be then more explicitly written as
\begin{align}
(\Lambda \circ \Gamma)[\eta]({\bf x}) - (\Gamma^T \circ \Lambda^T)[\eta]({\bf x}) &= \int d{\bf z}~d{\bf y}~\left[ \lambda({\bf x},{\bf z})~\gamma({\bf z},{\bf y}) - \gamma({\bf x},{\bf z}) \lambda({\bf z},{\bf y})\right]~\eta({\bf y}) \nonumber \\ &=\int~d{\bf y}~\left[ h({\bf x},{\bf y}) - h({\bf y},{\bf x})\right]~\eta({\bf y})=0,
\label{eq:comm-explicit}
\end{align}
where we have introduced the function
\begin{align}
h({\bf x},{\bf y}) = \int d{\bf z}~\lambda({\bf x},{\bf z})~\gamma({\bf z},{\bf y}).
\label{eq:hsmall}
\end{align}
If one wishes to have thermodynamic equilibrium, the expression in the second line of Eq.~\eqref{eq:comm-explicit} must be zero independently from the choice of $\eta({\bf y})$, which in turn imply the symmetry under the exchange of the arguments for the function $h({\bf x},{\bf y})$. The need of this symmetry for $h({\bf x},{\bf y})$ tells us that the symmetry under the exchange of arguments of the synaptic kernel $w({\bf x},{\bf y})$ is not in itself a sufficient condition to guarantee thermal equilibrium, what is needed is the symmetry of $h({\bf x},{\bf y})$, which implies a non trivial relation between $\gamma({\bf x},{\bf
  y})$ and $\lambda({\bf x},{\bf y})$. 
It is only in the case of a white noise which is delta-correlated also across space, i.e., $\gamma({\bf x},{\bf y})= T ~\delta({\bf x}-{\bf y})$, that symmetry of the synaptic kernel becomes sufficient to have equilibrium, since we have 
\begin{align}
h({\bf x},{\bf y}) = T~\gamma({\bf x},{\bf y}).
\end{align}
As shown in~\cite{lucente2022inference,lucente2025conceptual}, in the finite dimensional case these relations are equivalent to the Onsager reciprocal relations~\cite{PhysRev.37.405}. At the end of this section a comment is in order: since anyone is more familiar with the concept of entropy, rather than entropy production, it might be useful to show how the same expression of the entropy production obtained from a Lebowitz-Spohn approach connects to the variation in time of one of the possible definition of entropy in our system, namely the Shannon entropy. This task is accomplished in the next subsection, showing how an expression identical to the one of Eq.~\eqref{eq:sdot-general_final_trace} can be obtained. We deem this sort of derivation quite useful to grasp a more profound intuition of what entropy production is, before moving to Sec.~\ref{sec:EntropyFourier} where a more explicit expression is provided for the case of translational invariance. \\ \\

\section{Entropy production from Shannon entropy}
\label{sec:Shannon}

As anticipated in the discussion above, it is quite instructive to
explicitly derive the relation between the entropy production defined
within the Lebowitz-Spohn approach, specifically conceived for
stochastic systems, and the usual entropy considered in the canonical
ensemble, namely the Shannon entropy, which, according to the
prescription of stochastic
thermodynamics~\cite{seifert2005entropy,Sekimoto2010,seifert2012stochastic},
can be directly computed from the time-dependent functional
probability distribution $P_t[\eta]$ of the field $\eta({\bf x},t)$ as
\begin{equation}
    S_{\text{sys}}(t)=-\int \mathcal{D}\eta(t)~ P_t[\eta] \log\left(P_t[\eta]\right)\,,
    \label{eq:Shannon-entropy}
\end{equation}
where the symbol $\mathcal{D}\eta(t) = \prod_{\bf x}d\eta({\bf x},t)$
denotes functional integration over space at a given time $t$. In the
case of thermodynamic equilibrium, when the probability distribution
of the field $\eta({\bf x},t)$ does not depend on time and has the
form of a Boltzmann weight, for instance the one of
Eq.~\eqref{eq:stationary-equilibrium-pdf} in
Sec.~\ref{sec:stoch-amari-eq}, the Shannon entropy turns out to be
precisely the canonical entropy of the system, i.e., the one computed
as $S=\beta \langle E \rangle-\beta F$, where $\langle E \rangle$ and
$F$ are respectively the internal energy and the free energy at the
inverse temperature $\beta$. Also, interpreting $P_t$ as the single
particle marginal distribution of a $N-$dimensional system, $S_{sys}$
coincides with minus the $H-$function defined by
Boltzmann~\cite{de2013non}. But here we deal with a more general
case. We will show how, by taking the time derivative of the entropy
$S_{\text{sys}}(t)$ defined above and making use of the Fokker-Planck
equation, one ends up with terms among which it is possible to
recognize precisely the same entropy production expression derived
from the Lebowitz-Spohn approach, which we obtained in the previous
section. This sort of derivation is quite general, see for
instance~\cite{peliti2021stochastic,sarracino2025nonequilibrium}. Nevertheless,
in most places the procedure is discussed for the probability
distribution of a random variable, not of a fluctuating field, so that
we deemed helpful and not redundant to go through all steps of the
derivation using a {\it functional} formalism, which is what needed
here, and considering the specific form taken by the Fokker-Planck
equation in the presence of a functional formulation, which is not so
common in the stochastic thermodynamic literature. As we have said, we
are interested in the time derivative of $S_{\text{sys}}(t)$, which
reads as
\begin{align}
    \dot{S}_{\text{sys}}(t)&=-\int \mathcal{D}\eta(t) ~\dot{P}[\eta]\left[\log\left(P_t[\eta]\right)+1\right]~,\nonumber\\
     &= -\int \mathcal{D}\eta(t) ~\dot{P}[\eta]\log\left(P_t[\eta]\right).
     \label{eq:shannon-dot}
\end{align}
The second term added in the first line of Eq.~\eqref{eq:shannon-dot}
has been eliminated by exploiting the conservation of probability
normalization
\begin{align}
\int \mathcal{D}\eta(t)~\frac{d}{dt}P_t[\eta]= \frac{d}{dt} \left[\int\mathcal{D}\eta(t) ~P_t[\eta]\right]=0.
\end{align}
At this stage we need to use the Fokker-Planck which governs the dynamics of $P_t[\eta({\bf x},t)]$, which we write here in the form of a functional continuity equation which relates the rate of change probability density to the negative divergence of a probability current (or flux), thus ensuring the total probability conservation~\cite{o2025geometric}:
\begin{align}
    \frac{d}{dt}P_t[\eta]&=-\int {\rm d}{\bf x}\, \frac{\delta J[\eta]({\bf x})}{\delta \eta({\bf x},t)}\,,    \label{eq:fp}
\end{align}
where the current is defined as 
\begin{align}
    J[\eta]({\bf x}) &= P_t[\eta]\int {\rm d}{\bf y}\left[ \lambda({\bf x},{\bf y})\eta({\bf y},t)-\frac{1}{2}\gamma({\bf x},{\bf y})\frac{\delta}{\delta \eta({\bf y},t)}\log(P[\eta]) \right]
    \label{eq:fp-current}
\end{align}
Then, by inserting the time derivative of the probability distribution as written in Eq.~\eqref{eq:fp} into the expression of the Shannon entropy time-derivative, Eq.~\eqref{eq:shannon-dot}, one gets 
\begin{align}
    \dot{S}_{\text{sys}}(t)=\int \mathcal{D}\eta(t) \left[\int {\rm d}{\bf x}\, \frac{\delta J[\eta]({\bf x})}{\delta \eta({\bf x},t)}\right]    \log\left(P_t[\eta]\right)\,.
\end{align}
By integrating by parts the functional integral, one obtains\\
\begin{align}
 \dot{S}_{\text{sys}}(t) =- \int \mathcal{D}\eta~\int {\rm d}{\bf x}\,J[\eta]({\bf x})\frac{\delta}{\delta \eta({\bf x},t)}\log\left(P_t[\eta]\right)\,,
 \label{eq:shannon-dot-explicit}
\end{align}
where we have assumed that the probability current and probability
distribution vanish sufficiently rapid at the boundaries, so that the
boundary contribution can be neglected.  Let us notice that the above
formulation is completely general, it does not depend neither on the
linearity of the model nor on any stationarity assumption. If one then
takes into account the explicit expression of the current
$J[\eta]({\bf x})$ given in Eq.~\eqref{eq:fp-current}, and explicitly
substitutes $\frac{\delta}{\delta \eta({\bf
    x},t)}\log\left(P_t[\eta]\right)$ with
\begin{equation}
\frac{\delta}{\delta \eta({\bf x},t)}\log\left(P_t[\eta]\right)=2\int{\rm d}\mathbf{z}\,\gamma^{-1}(\mathbf{x},\mathbf{z})\left\{\int{\rm d}\mathbf{z'}\, \lambda({\bf z},{\bf z'})\eta({\bf z'},t)-\frac{J[\eta]({\bf z})}{P_t[\eta]}\right\}\,,
\end{equation}
it is possible to show that the time-derivative of the Shannon entropy
written in Eq.~\eqref{eq:shannon-dot-explicit} splits into the sum of
only two nontrivial contributions. These contributions, according to
the standard conventions in stochastic
thermodynamics~\cite{seifert2005entropy,seifert2012stochastic}, can be
identified respectively with the total entropy $S_{\text{tot}}(t)$ of
the universe (system+reservoir) and with the entropy of the reservoir
$S_{\text{res}}(t)$:
\begin{align}
    \dot{S}_{\text{sys}}(t)&= 
    \dot{S}_{\text{tot}}(t) - \dot{S}_{\text{res}}(t)
    \label{eq:tot-entr-var}
\end{align}
where we have
\begin{align}
    &\dot{S}_{\text{tot}}(t)=2\int \mathcal{D}[\eta]\int {\rm d}{\bf x} {\rm d}{\bf z} \frac{J[\eta]({\bf x}) \gamma^{-1}({\bf x},{\bf z})J[\eta]({\bf z})}{P_t[\eta]}\,,\label{eq:entropy-total}\\
    &\dot{S}_{res}(t)=2\int \mathcal{D}[\eta]\int {\rm d}{\bf x} {\rm d}{\bf z} {\rm d}{\bf z'}\, J[\eta]({\bf x})   \gamma^{-1}({\bf x},{\bf z})\lambda({\bf z},{\bf z'})\eta({\bf z'})\label{eq:entropy-medium}\, . 
\end{align}
The relation between the rate of change of the Shannon entropy and the Lebowitz-Spohn approach emerges in the stationary state, where  
\begin{align}
\frac{d}{dt} P[\eta] = 0 ~~~\Longrightarrow~~~ \dot{S}_{\text{sys}}=0~~~\Longrightarrow~~~\dot{S}_{\text{tot}}=\dot{S}_{\text{res}}
\label{eq:entropy-implications}
\end{align}
Physically, the last identity of Eq.~\eqref{eq:entropy-implications} means that in the stationary state the totality of the entropy produced during the dynamics is dissipated into the environment/reservoir. The last step of the calculation amounts to show that, in the stationary state, this rate of change of the entropy dissipated by the environment corresponds precisely to the entropy production rate according to the Lebowitz-Spohn formula, i.e. that we have precisely
\begin{align}
\sigma = \dot{S}_{\text{tot}}=\dot{S}_{\text{res}}
\end{align}
In order to show this, we need to explicitly write the expression of $P[\eta]$, which 
in the stationary state is Gaussian:
\begin{align}
P[\eta] \propto \exp\left\lbrace - \frac{1}{2}\int d{\bf x}d{\bf y}~\eta({\bf x})c^{-1}({\bf x},{\bf y})\eta({\bf y})\right\rbrace\,,
\end{align}
where the kernel is the inverse of the stationary field correlator $c({\bf x},{\bf y}) = \langle \eta({\bf x}) \eta({\bf y}) \rangle$. Then, by plugging this stationary probability into the expression of the current one gets
\begin{align}
    J[\eta]({\bf x}) &= P[\eta]\int {\rm d}{\bf y}\left[ \lambda({\bf x},{\bf y})\eta({\bf y})+\frac{1}{2}\int {\rm d}{\bf z} \gamma({\bf x},{\bf y})c^{-1}({\bf y},{\bf z})\eta({\bf z}) \right]\,.
    \label{eq:fp-current-linear}
\end{align}
Finally, by inserting the expression of the current Eq.~(\ref{eq:fp-current-linear}) into the expression of the reservoir entropy derivative $\dot{S}_{\text{res}}(t)$, and by exploiting the same operator identities of Sec.~\ref{sec:entropy-prod-gen}, one gets
\begin{align}
    \dot{S}_{\text{res}}&=2 \int {\rm d}{\bf x} {\rm d}{\bf z} {\rm d}{\bf z'}{\rm d}{\bf y}\, \big[\lambda({\bf x},{\bf y})   \gamma^{-1}({\bf x},{\bf z})\lambda({\bf z},{\bf z'})\langle\eta({\bf z'})\eta({\bf y})\rangle 
    \nonumber \\
    &+\frac{1}{2}\int {\rm d}{\bf q} \gamma({\bf x},{\bf y})c^{-1}({\bf y},{\bf q}) \gamma^{-1}({\bf x},{\bf z})\lambda({\bf z},{\bf z'})\langle\eta({\bf z'})\eta({\bf q})\rangle) \big]\nonumber\\
    &=\text{Tr}\left[2\Lambda^T \circ \Gamma^{-1}\circ \Lambda \circ C+\Lambda\right]
    \nonumber \\ 
    &=\text{Tr}\left[\Lambda^T \circ \Gamma^{-1}\circ \Lambda \circ C\right]+\text{Tr}\left[-\Lambda^T\circ \Gamma^{-1}\circ (C\circ \Lambda^T+\Gamma)+\Lambda\right]\nonumber\\
    &= \text{Tr}\left[(\Lambda^T \circ \Gamma^{-1}-\Gamma^{-1}\circ\Lambda)\circ \Lambda \circ C\right]\,.
    \label{eq:epr_medium}
\end{align} 
The above formula coincides exactly with
Eq.\eqref{eq:sdot-general_final_trace}. We have therefore shown an
alternative way to derive the same expression obtained from the
Lebowitz-Spohn approach.

\section{Entropy production in the bulk: an explicit formula}
\label{sec:EntropyFourier}

In order to provide more explicit expressions of the entropy
production rate a convenient and still quite general assumption which
simplifies a lot the calculation, allowing for final explicit
expression, is to claim translational invariance in space for the
synaptic weight function. The assumption of translational invariance
clearly does not fit the brain, if we think about it as a whole, since
it does not have infinite extension neither is a periodic system. It
has well defined boundaries and a specific geometric
shape. Nevertheless, if we imagine to focus on the properties of a
small piece in the bulk of the brain, then, as is it customary in the
statistical mechanics approach to condensed matter, it then makes
sense to assume periodic boundary conditions and, as a consequence of
that, translational invariance. In practice this corresponds to assume
that the synaptic weight kernel depends only on the difference between
the two coordinates:
\begin{align}
w({\bf x},{\bf y}) = w({\bf x}-{\bf y})
\end{align}
By exploiting the translational invariance of the synaptic weight
function it is possible to consider the Fourier transform of the
Langevin equation for the field fluctuations,
Eq.~\eqref{eq:Ama-compact}, so to have $M$ uncoupled equations, where
$M$ is the number of Fourier modes, ${\bf k} \in [{\bf
    k}_{0},\ldots,{\bf k}_M]$, of the kind:
\beq
\frac{\partial}{\partial t} \eta_{\bf k}(t) = \lambda_{\bf k}\, \eta_{\bf k}(t) + \xi_{\bf k}(t)
\;
 \label{eq:compact-laf}
\eeq
where
\beq
\lambda_{\bf k} = - \frac{1}{\tau} + \frac{f'[u_0]}{\tau} \, w_{\bf k} \;  
\label{eq:lambdak}
\eeq
and where we have introduced the Fourier transform of the
synaptic weight
\begin{align}
w_{\bf k}= \frac{1}{(2\pi)^{d/2}} \int d{\bf x}~e^{i {\bf k}{\bf x}}~w({\bf x}).
\end{align}
\bcol{At variance with the synaptic kernel $w({\bf x},{\bf y})$, the assumption of translational invariance for the noise kernel is quite natural, namely we have $\gamma({\bf x},{\bf y})=\gamma({\bf x}-{\bf y})$. Since the noise correlator is then, by definition, symmetric, it follows that its Fourier transform $\gamma_{\bf k}$ is a real
function, i.e., $\gamma_{\bf k}=\text{Re}[\gamma_{\bf k}]$.} Let us now specify, given the forward trajectory $\eta_{\bf k}(s)$ with $s\in [0,t]$, the definition of its corresponding backward trajectory $\overline{\eta_{\bf k}(s)}$, which, consistently with its definition in real space, is defined as:
\begin{align}
\overline{\eta_{\bf k}(s)}&=
\eta_{\bf k}(t-s),
\end{align}
and also
\begin{align}
\frac{d\overline{\eta_{\bf k}(s)}}{d s} &= -\frac{d}{du}\eta_{\bf k}\Big\vert_{u=(t-s)}.
\end{align}
From the above definition we have therefore that, if the forward trajectory is a solution of the Langevin equation in Eq.~\eqref{eq:compact-laf}, the backward trajectory is then a solution of
\begin{align}
\frac{d}{du} \eta_{\bf k}\Big\vert_{u=(t-s)} =
-\lambda_{\bf k} \, \eta_{\bf k}(t-s) - \xi_{\bf k}(t-s)
\end{align}

In terms of the Fourier modes of the field the probabilities of
forward and backward trajectories can be rewritten, using a compact
notation for time derivatives, $\dot{\eta}_{\bf
  k}=\frac{d}{du}\eta_{\bf k}$, as:

\begin{align}
\mathcal{P}[\Omega_0^t] &\propto \exp\left\lbrace - \frac{1}{2} \int_0^t ds
\sum_{\bf k}~ \left[\dot{\eta}^*_{\bf k}(s)-\lambda^*_{\bf k}\eta^*_{\bf k}(s)\right]\gamma^{-1}_{\bf k}\left[ \dot{\eta}_{\bf k}(s)-\lambda_{\bf k}\eta_{\bf k}(s) \right] \right\rbrace, \nonumber \\
\mathcal{P}[\overline{\Omega}_0^t] &\propto \exp\left\lbrace  - \frac{1}{2} \int_0^t ds
\sum_{\bf k}~ \left[\dot{\eta}^*_{\bf k}(t-s)+\lambda_{\bf k}^*\eta^*_{\bf k}(t-s)\right]\gamma^{-1}_{\bf k}\left[ \dot{\eta}({\bf k},t-s)+\lambda_{\bf k}\eta_{\bf k}(t-s) \right] \right\rbrace,
\label{eq:prob-traj-k}
\end{align}
 which, exploiting the mathematical identity
\begin{align}
  \int_0^t ds~f(t-s) &= \int_0^t ds~f(s) 
\end{align}
can be rewritten as
\begin{align}
\mathcal{P}[\Omega_0^t] &\propto \exp\left\lbrace - \frac{1}{2} \int_0^t ds
\sum_{\bf k}~ \frac{|\dot{\eta}_{\bf k}(s)|^2+|\lambda_{\bf k}\eta_{\bf k}(s)|^2-\dot{\eta}^*_{\bf k}(s) \lambda_{\bf k}\eta_{\bf k}(s) - \dot{\eta}_{\bf k}(s) \lambda^*_{\bf k}\eta^*_{\bf k}(s)}{\gamma_{\bf k}} \right\rbrace,  \\
\mathcal{P}[\overline{\Omega}_0^t] &\propto \exp\left\lbrace  - \frac{1}{2} \int_0^t ds
\sum_{\bf k}~ \frac{ |\dot{\eta}_{\bf k}(s)|^2+|\lambda_{\bf k}\eta_{\bf k}|^2(s)+\dot{\eta}^*_{\bf k}(s) \lambda_{\bf k}\eta_{\bf k}(s)+
\dot{\eta}_{\bf k}(s)\lambda^*_{\bf k}\eta^*_{\bf k}(s)}{\gamma_{\bf k}}\right\rbrace.
\label{eq:prob-traj-k}
\end{align}

Let us notice that in the above Eq.~\eqref{eq:prob-traj-k} the field
$\eta_{\bf k}(s)$ and $\lambda_{\bf k}$ take values in
$\mathbb{C}$, since there is no general reason to assume that $w({\bf
  x})$ has a definite parity symmetry. Then, according to its definition, the entropy production rate for
the stochastic Amari model with translational invariant synaptic
weight reads as
\begin{align}
  \bcol{\sigma} = \lim_{t\rightarrow\infty}\frac{2}{t}\int_0^t ds~\sum_{\bf k} \frac{\text{Re}\left[\lambda_{\bf k}\dot{\eta}_{\bf k}^*(s)\eta_{\bf k}(s)\right]}{\gamma_{\bf k}} = \frac{\lambda_{\bf k}}{\gamma_{\bf k}} \langle \dot{\eta}^*_{\bf k}\eta_{\bf k}\rangle +   \frac{\lambda^*_{\bf k}}{\gamma_{\bf k}} \langle \dot{\eta}_{\bf k}\eta^*_{\bf k}\rangle
  \label{eq:sdot-k1}
\end{align}  

Further progresses in the explicit calculation of the entropy production rate per Fourier mode \bcol{$\sigma_{\bf k}$} can be done
by considering explicitly the formal solution of the Langevin
Eq.~\eqref{eq:compact-laf}:
\begin{align}
\eta_{\bf k}(t) = e^{\lambda_{\bf k}t}~\eta_{\bf k}(0) + \int_0^t ds~e^{\lambda_{\bf k}(t-s)}~\xi_{\bf k}(s), 
\end{align}
from which, under the not too restrictive assumption $\eta_{\bf
  k}(0)=0$, one immediately gets
\begin{align}
  \eta_{\bf k}(t) &=\int_0^t ds~e^{\lambda_{\bf k}(t-s)}~\xi_{\bf k}(s) \\
  \dot{\eta}_{\bf k}(t) &= \xi_{\bf k}(t) + \lambda_{\bf k}\int_0^t ds~e^{\lambda_{\bf k}(t-s)} \xi_{\bf k}(s)
\end{align}
From the above expressions we can compute for instance $\langle \dot{\hat{\eta}}({\bf k},t)\hat{\eta}^*({\bf k},t)\rangle$:
\begin{align}
  \langle \dot{\eta}_{\bf k}(t)\eta^*_{\bf k}(t)\rangle &= \int_0^t d\tilde{s}~e^{\lambda^*_{\bf k}(t-\tilde{s})}
  \langle \xi_{\bf k}(t)\hat{\xi}^*_{\bf k}(\tilde{s})\rangle + \lambda_{\bf k}
  \int_0^t d\tilde{s}~ds~e^{\lambda^*_{\bf k}(t-\tilde{s})+\lambda_{\bf k}(t-s)}~\langle\xi_{\bf k}(s)\xi^*_{\bf k}(\tilde{s})\rangle \nonumber \\
  & = \gamma_{\bf k}~\left[ \frac{1}{2} + \lambda_{\bf k} \int_0^t ds~e^{[\lambda^*_{\bf k}+\lambda_{\bf k}](t-s)}\right] \nonumber \\
  & = \gamma_{\bf k}~\left[ \frac{1}{2} + \frac{\lambda_{\bf k}}{\lambda^*_{\bf k}+\lambda_{\bf k}} e^{2\text{Re}[\lambda_{\bf k}]t} - \frac{\lambda_{\bf k}}{\lambda^*_{\bf k}+\lambda_{\bf k}} \right]. \nonumber \\
\end{align}
Since we are interested in the value of the above correlation function at stationarity, we send $t\rightarrow\infty$ and,
assuming $\text{Re}[\lambda_{\bf k}]<0$, we have
\begin{align}
  \langle \dot{\eta}_{\bf k}\eta^*_{\bf k}\rangle = \lim_{t\rightarrow\infty} \langle \dot{\eta}_{\bf k}(t)\eta^*_{\bf k}(t)\rangle
  =
\frac{\gamma_{\bf k}}{2\text{Re}[\lambda_{\bf k}]}\left[ \frac{\lambda^*_{\bf k}-\lambda_{\bf k}}{2}\right].
\end{align}
Correspondingly, it is straightforward to have that the complex
conjugate of the same quantity reads as:

\begin{align}
\langle \dot{\eta}^*_{\bf k}\eta_{\bf k}\rangle = \lim_{t\rightarrow\infty} \langle \dot{\eta}^*_{\bf k}(t)\eta_{\bf k}(t)\rangle =
\frac{\gamma_{\bf k}}{2\text{Re}[\lambda_{\bf k}]}\left[ \frac{\lambda_{\bf k}-\lambda^*_{\bf k}}{2}\right].
\end{align}

By plugging the above results on stationary correlators into the
expression of entropy production in Eq.~\eqref{eq:sdot-k1} one gets:

\begin{align}
  \sigma_{\bf k} =  \frac{\lambda_{\bf k}}{\gamma_{\bf k}} \langle \dot{\eta}^*_{\bf k}\eta_{\bf k}\rangle +   \frac{\lambda^*_{\bf k}}{\gamma_{\bf k}} \langle \dot{\eta}_{\bf k}\eta^*_{\bf k}\rangle 
  = i\,\,\text{Im}[\lambda_{\bf k}]\frac{(\lambda_{\bf k}-\lambda^*_{\bf k})}{2\text{Re}[\lambda_{\bf k}]}
\end{align}
and consequently 
\begin{equation}
\sigma_{\bf k} = - \frac{\text{Im}^2[\lambda_{\bf k}]}{\text{Re}[\lambda_{\bf k}]}.
\end{equation}
This leads us to the final result for the entropy production
\begin{equation}
\sigma = - 
\sum_{\bf k} 
\frac{\text{Im}^2[\lambda_{\bf k}]}{\text{Re}[\lambda_{\bf k}]} \; ,  
\end{equation}

which is a remarkably simple formula in terms of the eigenvalues
$\lambda_{\bf k}$ of the linearized Amari equation in Fourier space.

\section{Conclusions}
\label{sec:conclusion}

In this work, we have derived an explicit and compact expression for
the entropy production rate $\sigma$ in the stochastic, linearized
Amari neural field model, thereby providing a rigorous bridge between
nonequilibrium statistical mechanics and large-scale neural
dynamics. By computing $\sigma$, we have shown that the emergence of
irreversibility in the stochastic Amari model is entirely determined
by the interplay between the symmetry of the synaptic kernel
$w(\mathbf{x},\mathbf{y})$ and the structure of the noise covariance
$\gamma(\mathbf{x},\mathbf{y})$. Moreover, we have discussed the
relation between the entropy production rate $\sigma$, as defined by
Lebowitz and Spohn, and the temporal variation of the system entropy
$\dot{S}_{\text{sys}}$, showing that, in a nonequilibrium stationary
state, $\sigma$ equals the entropy dissipated into the reservoir
$\dot{S}_{\text{res}}$. While this equivalence is well known in
stochastic
thermodynamics~\cite{Sekimoto2010,seifert2005entropy,peliti2021stochastic},
we derived it by employing the functional formalism typical of neural
field models. Thus, our derivation systematically connects functional
formulations of stochastic neural-field
theory~\cite{wilson1972excitatory,bressloff2011spatiotemporal} with
the language of modern stochastic
thermodynamics~\cite{peliti2021stochastic}.  Most previous literature
on the subject has aimed at developing methods for estimating the
entropy production rate or other measures of irreversibility from
experimental signals and, therefore, usually deals with
finite-dimensional
models~\cite{lynn2021broken,PhysRevE.107.024121,PhysRevResearch.7.013301,PhysRevE.110.034313}. In
contrast, we have carefully examined the mechanism of time-reversal
symmetry breaking in the idealized framework of the stochastic Amari
model. It is worth emphasizing that this model lacks several
biological ingredients, such as population heterogeneity, delays in
signal propagation and synaptic plasticity, since it describes a
single scalar field $\eta(\mathbf{x},t)$ and employs a
time-independent coupling kernel $w(\mathbf{x},\mathbf{y})$. Moreover,
it is a phenomenological model, meaning that it is not derived from a
microscopic description through a large-scale expansion. As explained
in Sec.~\ref{sec:stoch-amari-eq}, the fluctuations are externally
imposed on the deterministic dynamics. To discuss, in the simplest
way, the interplay between synaptic coupling and noise correlations,
while avoiding certain interpretational subtleties, we focused on the
case of additive Gaussian noise. Within this approximation we are
neglecting many possible sources of irreversibility such as those
produced by non-Gaussian, non-Markovian or multiplicative
fluctuations. Despite these limitations, our analysis clarifies the
conditions that $w(\mathbf{x},\mathbf{y})$ and
$\gamma(\mathbf{x},\mathbf{y})$ must satisfy for linear, spatially
extended neural fields to not produce entropy during the
evolution. For instance, the analysis in Fourier space, when
translational invariance was assumed, revealed that spatial
correlations of fluctuations do not contribute to the time-reversal
symmetry breaking, which originate solely from asymmetries in the
coupling kernel. In this case, we have also shown that the entropy
production rate $\sigma$ can be decomposed into positive contributions
$\sigma_\mathbf{k}$, one for each mode $\mathbf{k}$. This
decomposition, similar to that proposed
by~\cite{sekizawa2024decomposing}, but applied in wave-vector space
$\mathbf{k}$ rather than in frequency domain, provides a simple
theoretical tool to probe dissipation in neuronal systems across
different spatial scales associated with the Fourier modes. In
conclusion, the simplified setting adopted here should be understood
as a testbed for uncovering the mechanisms by which asymmetries in
synaptic couplings and noise spatial correlations generate
irreversible steady dynamics. Isolating these contributions to entropy
production may prove useful when more realistic situations are
considered. Moreover, it has allowed us to establish the theoretical
foundation for linking stochastic thermodynamics to spatially extended
neural-field models under minimal assumptions and can serve as a
starting point for future explorations of energy dissipation and
information flow in neuronal systems.

\acknowledgments{G.G. acknowledge partial support from the project
  MIUR-PRIN2022, ``Emergent Dynamical Patterns of Disordered Systems
  with Applications to Natural Communities'', code
  2022WPHMXK. D.L. acknoledge support from NEXTGENERATIONEU (NGEU)
  funded by the Ministry of University and Research (MUR), National
  Recovery and Resilience Plan (NRRP), and Project MNESYS
  (PE0000006)-A multiscale integrated approach to the study of the
  nervous system in health and disease (DN. 1553
  11.10.2022). L.L. acknowledge partial support from
  ``IniziativaSpecifica Quantum'' of INFN, by the European
  Union-NextGenerationEU within the National Center for HPC, Big Data
  and Quantum Computing (Project No. CN00000013, CN1 Spoke 10:
  ``Quantum Computing''), by the EU Project PASQuanS 2 ``Programmable
  Atomic Large-Scale Quantum Simulation'', and the National Grant of
  the Italian Ministry of University and Research for the PRIN 2022
  project ``Quantum Atomic Mixtures: Droplets, Topological Structures,
  and Vortices''.}

\bibliography{bibliography}

\end{document}